\documentclass[prd,aps,12pt]{revtex4}
\usepackage{verbatim}
\usepackage{color}
\usepackage{rotate}
\usepackage[dvips]{graphics}
\usepackage{epsf}
\usepackage{epsfig}
\usepackage{graphicx}
\usepackage{latexsym}
\usepackage{amscd}
\usepackage{amssymb}
\newcommand{\dens}{{\widehat{w}}}
\newcommand{\wh}{\widehat}
\newcommand{\ul}{\underline}
\renewcommand{\ol}{\overline}
\newcommand{\beq}{\begin{equation}}
\newcommand{\eeq}{\end{equation}}
\newcommand{\beqn}{\begin{eqnarray}}
\newcommand{\eeqn}{\end{eqnarray}}
\newcommand\noi{\noindent}
\newcommand\la{\langle}
\newcommand\ra{\rangle}
\newcommand\eps\varepsilon
\newcommand\euler{{\rm e}}
\newcommand\imag{{\imath}}

\newcommand\sgn{{\rm sgn}}

\newcommand\trans{{\wh{\cal P}}}

\def\lsim{\mathrel{\rlap{\lower4pt\hbox{\hskip1pt$\sim$}}
    \raise1pt\hbox{$<$}}}
\def\gsim{\mathrel{\rlap{\lower4pt\hbox{\hskip1pt$\sim$}}
    \raise1pt\hbox{$>$}}}

\makeatletter
\def\fmslash{\@ifnextchar[{\fmsl@sh}{\fmsl@sh[0mu]}}
\def\fmsl@sh[#1]#2{%
  \mathchoice
    {\@fmsl@sh\displaystyle{#1}{#2}}%
    {\@fmsl@sh\textstyle{#1}{#2}}%
    {\@fmsl@sh\scriptstyle{#1}{#2}}%
    {\@fmsl@sh\scriptscriptstyle{#1}{#2}}}
\def\@fmsl@sh#1#2#3{\m@th\ooalign{$\hfil#1\mkern#2/\hfil$\crcr$#1#3$}}
\makeatother

\begin{document}

\hfill {SLAC-PUB-10605}


\title{Statistical Physics and Light-Front Quantization}

\author{J\"org Raufeisen\footnote{\tt email: jorg@slac.stanford.edu} and
Stanley J. Brodsky\footnote{\tt email: sjbth@slac.stanford.edu}
}
\medskip

\affiliation{Stanford Linear Accelerator Center, Stanford University,
2575 Sand Hill Road, Menlo Park, CA 94025, USA}

\vspace{2cm}


\begin{abstract}
\vspace*{1cm} 
\centerline{\bf Abstract} \noi
Light-front quantization has important advantages for describing relativistic
statistical systems,
particularly
systems for which boost invariance
is essential, such as the fireball created in a heavy ion collisions.
In this paper we develop light-front field theory at finite
temperature and density with special attention to  quantum chromodynamics.
We construct the most general form of the statistical
operator allowed by  the Poincar{\'e} algebra and 
show that there are no zero-mode related problems when
describing phase transitions.
We then demonstrate a direct connection between densities in
light-front thermal field theory and the parton distributions measured in
hard scattering
experiments. Our approach thus generalizes the concept of a parton
distribution
to finite temperature. In light-front quantization, the
gauge-invariant Green's functions of a quark in a medium can be defined in
terms of just
2-component spinors and have a much simpler spinor structure than the 
equal-time fermion propagator. From the
Green's function, we introduce the new concept of a light-front density matrix,
whose matrix elements are related to forward and to
off-diagonal
parton distributions.   Furthermore, we explain how thermodynamic
quantities can be
calculated in discretized light-cone quantization, which is  applicable at 
high chemical potential and is not plagued
by the fermion-doubling problem.

\noi
\end{abstract}
\maketitle


\section{Introduction}
\label{sec:Introduction}

Dirac's front form of relativistic dynamics \cite{Dirac} has remarkable
advantages for relativistic problems in high energy and nuclear physics.
Most appealing is the
simplicity of the vacuum (the ground state of the free theory is also the
ground state of the
full theory) and the existence of boost-invariant light cone wavefunctions (see
Ref.~\cite{BPPreport} for a review.) This makes light-front quantization a natural candidate for the description of systems for which boost
invariance is an issue,
such as the fireball created in a heavy ion collision or the small-$x$
features  of a nuclear
wavefunction.    It is clearly important to exploit
the advantages of light-front quantization 
for thermal field theory \cite{diplom,Stan1}.
Valuable work in
this direction  has already been done by several authors \cite{alex,Beyer2,Alves,
Weldon, Weldon2, Das,
Blankleider, Beyer}.  In this paper we shall  apply light-front quantization to
statistical physics and investigate the prospects and challenges of this
approach for
quantum chromodynamic systems.

\begin{figure}[t]
  \centerline{\scalebox{1.0}{\includegraphics{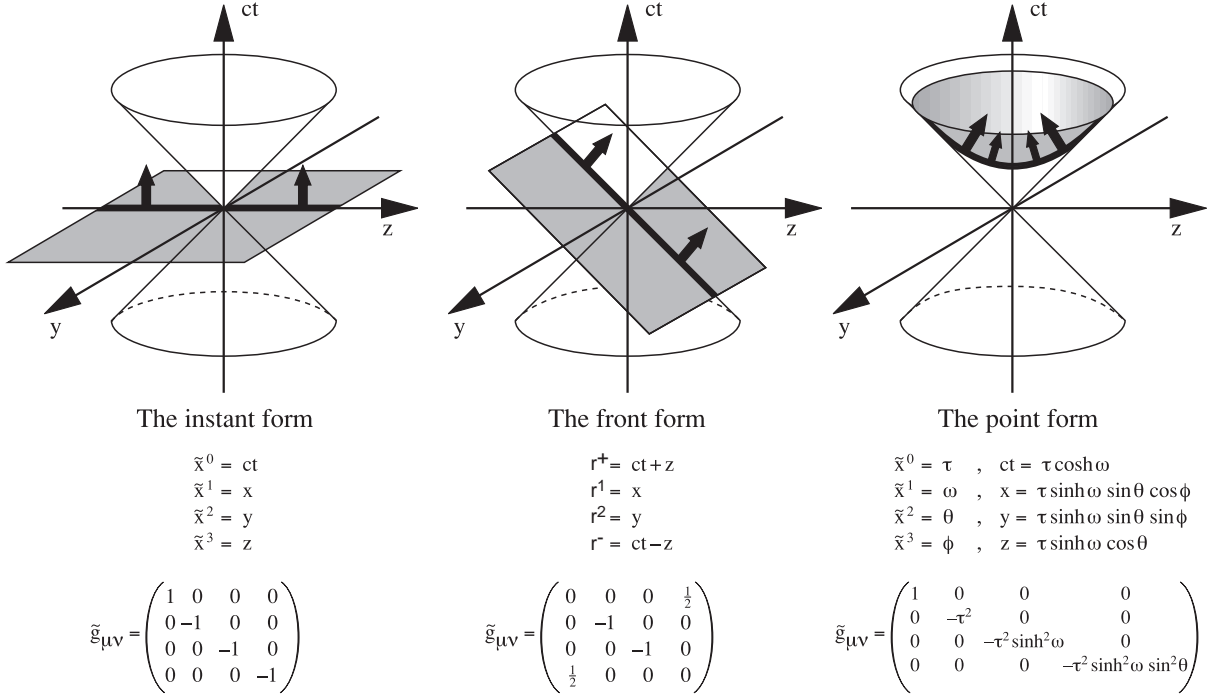}}}
    \center{
    {\caption{\em
      \label{fig:dirac}The three forms of Hamiltonian dynamics \cite{Dirac,BPPreport}.  In the
front form (middle), initial conditions are defined on a light-like
surface, and the
system is the propagated in light-cone time $x^+$ by the light-cone
Hamiltonian $\wh
P^-$.  }
    }  }
\end{figure}

In the front form, see Fig.~\ref{fig:dirac}, initial conditions are defined on a
light-like hypersurface with $r^+=r^0+r^3=0$.  (A summary of our notation
is given in
Appendix \ref{sec:conventions}.) This sets the boundary condition as the
light front
moves forward.  When the theory is quantized, (anti-~) commutation
relations are defined
at fixed light-front time $r^+$, instead of fixed equal-time $r^0$.  As a
consequence, the
number of kinematic Poincar{\'e} generators in the front-form is larger than 
in any other form
\cite{Dirac,
BPPreport}, namely 7 out of 10 Poincar{\'e} generators do not depend on the
interaction.
In particular, the generator of Lorentz boosts in the $r^3$ direction is a
kinematic
operator on the light front, but not in the instant form.  Therefore, the
statistical
operator in light-front quantization will transform trivially under
longitudinal boosts.

Another advantage of the light-front formulation of thermal field theory is
that the
charge is defined as the integral over the $+$-component of the current density,
$Q=\int d^3r
j^+(\vec r)$.  This allows one to establish a clear connection between
densities in
thermal field theory and the boost-invariant
parton distributions measured in hard
scattering experiments
\cite{fact}.  No such relation exists in the instant form, where charges are
integrals
over $0$-components of current densities. Also, parton distributions defined
in equal-time quantization are not boost invariant, since the generator of Lorentz 
transformations in $r_3$ direction contains interaction terms. 
As a consequence, in this formalism one might regard low-$x$ gluon saturation 
as a phase transition \cite{BNL} or a critical phenomenon \cite{Pirner}
in the sense of statistical mechanics.
One can even generalize the concept of a
parton
distribution to finite temperature, which is relevant for jet quenching in
heavy ion
collisions \cite{Ivan} as well as for parton recombination \cite{Rainer}.

We also point out that fermions behave completely differently on the
light-front, where
they are described terms of just 2-component spinors.  It turns out that
the fermion
propagator has only one derivative in the numerator, leading to only one
pair of doublers
on the lattice.  In addition, the doubler problem is entirely avoided in
Discretized
Light-Cone Quantization (DLCQ) \cite{DLCQ}.  This technique is also
applicable at finite
density and will hopefully allow one to study baryon-rich QCD matter in
large scale
numerical calculations.  Despite recent progress \cite{Fodor}, conventional
lattice QCD is
still notoriously complicated at finite baryon chemical potential.

This paper is organized as follows.
In the next section,  we
construct the most general form of the
statistical
operator allowed by  the Poincar{\'e} algebra.  Our results are in
agreement with earlier
findings \cite{diplom,Das,Weldon}.  We
argue that the
simplicity of the vacuum does not lead to difficulties when describing phase
transitions.  In Section \ref{sec:green}, we define gauge-invariant Green's
functions for
fermions at finite temperature and relate them to generalized parton distributions 
in Section \ref{sec:gpds}.
In the summary, we give an outlook on further applications and challenges
for this approach.

\section{The statistical operator at finite temperature and density}
\label{sec:construction}

The form of the statistical operator $\dens$ at finite temperature and density can be
obtained from very general considerations.  Our result below for $\dens$,
Eq.~(\ref{eq:general}), is compatible with the findings of \cite{Alves,
Weldon}, {\em
i.e.} $\dens$ is always the exponential of the equal time
energy $\wh P^0$
in the local rest frame of the system.  Here, we shall give a more physical
derivation of
that result, following  Ref.~\cite{Landau5}.

In the instant form, the equal time Liouville theorem
($\imag\partial^0\dens_{ET}=[\wh
P^0,\dens_{ET}]$) requires that in equilibrium, $\dens_{ET}$ is a
function of
only those Poincar{\'e} generators which commute with the equal time
Hamiltonian $\wh
P^0$.  In addition, since  systems far apart from each other must be
uncorrelated, the density operator of the combined system has to factorize
into the
density operators of the subsystems. 
This requirement is similar to the cluster decomposition principle that
has been studied in the context of the deuteron wavefunction
on the light-cone in Ref.~\cite{cluster}.
Consequently, in equilibrium
$\ln(\dens_{ET})$ must
be a linear combination of the additive constants of motion, namely the
four components
of the momentum $\wh P^\mu$ and the angular momentum three-vector $\vec J$.
Hence,
$\ln(\dens_{ET})=\alpha-\beta(u_\nu \wh P^\nu -\vec\omega\cdot\vec
J-\sum_l\mu_l\wh
Q_l)$, where $\alpha$ is a normalization constant.  The notation is chosen
in anticipation
of the physical meaning of the various coefficients: $\beta$ is the inverse
temperature
and $u_\nu$ is the four velocity of the system, cf.~Ref.~\cite{Alves}.
In addition,
$\vec\omega$ is the angular velocity at which the body rotates.  Additional
conserved
charges $\widehat Q_l$ are included along with their chemical potentials
$\mu_l$.  We
stress that the ``charges", {\em i.e.} the mean values of the operators
$\wh P_\nu,
\vec J, \wh Q_l$, are given and that the ``chemical potentials" ($u_\nu,
\vec\omega,
\mu_l$) need to be determined from the conservation laws.  In quantum
mechanics, of
course, one can simultaneously specify only charges which commute with each
other, {\em
i.e.} only one component of $\vec J$ (namely for rotations around the axis
with the
largest moment of inertia.) For the same reason, one cannot specify all
four $P_\nu$
for systems with nonzero angular momentum.

What changes, when we switch to the light front?  First of all, the equal
time Liouville
theorem has to be replaced by its light front analog.  Writing the light
front density
operator as \beq \dens=\sum_{ij}c_{ij}\left|\phi_i\ra\la\phi_j\right|, \eeq
the (light
cone) time ($x^+$) evolution of the multi particle state
$\left|\phi_i\ra\right.$ is
generated by the second quantized Hamiltonian $\wh P^-$, \beq
\imag\partial^-\left|\phi_i\ra\right.=\widehat P^-\left|\phi_i\ra\right..
\eeq From this,
the light front version of the Liouville theorem immediately follows:
\beq\label{eq:liouville} \imag\partial^-\dens=\left[\widehat
P^-,\dens\right]. \eeq
Equilibrium (on the light front) is reached, when the commutator on the
right hand side
({\em rhs}) of Eq.~(\ref{eq:liouville}) vanishes.  The six Poincar{\'e}
generators which
commute with $\widehat P^-$ are the other three momentum components and
three of the six
independent elements of the antisymmetric boost-angular momentum tensor
\beq\label{eq:mmunu} \widehat M^{\kappa\nu}= \left(
\begin{array}{cccc}
0 & \wh K_1+\wh J_2 & \wh K_2-\wh J_1 & -2\wh K_3 \\
-\wh K_1-\wh J_2 & 0 & \wh J_3 & -\wh K_1+\wh J_2 \\
-\wh K_2+\wh J_1 & -\wh J_3 & 0 & -\wh K_2-\wh J_1 \\
2\wh K_3 & \wh K_1-\wh J_2 & \wh K_2+\wh J_1 & 0
\end{array}\right)
\eeq ($\mu,\nu\in\{+,1,2,-\}$).  Namely the rotations ($\widehat
M^{1,2}=\widehat J_3$)
around the longitudinal direction and the two dynamic operators $\wh
M^{\perp-}=(\widehat
M^{1-},\widehat M^{2-})$, where $\widehat M^{1-}=-\widehat K_1+\widehat
J_2$, $\widehat
M^{2-}=-\widehat K_2-\widehat J_1$.  The $\wh K_j$ are boost generators in
$j$ direction.
Since two of the operators $\wh M^{\perp-}$ have zero eigenvalues when
acting on all
physical states, the light front density operator has the general form
\beq\label{eq:general} \ln(\dens)=\alpha-\beta\left(u_\nu \widehat
P^\nu-\omega \wh
J_3-\sum_l \mu_l\widehat Q_l\right). \eeq Thus we have obtained the same J\"uttner type
distribution as in the instant form.  The identification of $u_\nu$ with the
four-velocity of the system relative to the observer can be derived by
calculating the commutators of $\ln(\dens)$ with the boost generators by
means of the
Poincar{\'e} algebra. Since $u_\nu u^\nu=1$, only four of the five parameters 
$\beta=1/T$ and $u^\nu$ are independent. 

We remark that $T$ is the same as the 
instant-form temperature, but the chemical potential has a different meaning.  
On the light-front, densities are given by
$+$-components of currents, and not by $0$-components.  This is essential
for the proper
generalization of parton distributions to finite temperature, since the
latter are also
defined as $+$-components \cite{fact}.

Note that one can include rotations in the light front formalism, but one
has to make
sure that the longitudinal axis has the largest inertial moment. Of course,
one cannot specify $\vec P_\perp$ and $J_3$ simultaneously. We shall
not study
systems with finite angular momentum in what follows.  

We choose $\alpha=0$ as
normalization so that the partition function is given by ${\cal Z}={\rm
Tr}\dens$.  Since
$\cal Z$ is a Lorentz scalar, all thermodynamic potentials and the entropy
transform as
scalars, {\em e.g.} the Lorentz invariant generalization of the grand-canonical potential
(or of the free energy in the case of $\mu=0$) is
\beq
\Omega=-T\ln{\cal Z}(V,T,\mu),
\eeq
and the entropy is defined by
\beq
S=-\left(\frac{\partial\Omega}{\partial T}\right)_{\mu,V}=-\frac{1}{\cal Z}{\rm Tr}(\dens\ln\dens).
\eeq
The role of the total energy of the system is now played by the expectation value
of $u_\nu\wh P^\nu$,
\beq
U=\la u_\nu\wh P^\nu\ra=\frac{u_\nu}{\cal Z}{\rm Tr}(\dens \wh P^\nu).
\eeq
As usual, $\Omega=U-TS-\mu Q$. All known relations between thermodynamic potentials remain valid.

Even though one can formulate light-front statistical mechanics in an
arbitrary frame, it
is convenient to choose the reference frame in which the system as a whole
is at rest, so
that $\ln(\dens)=-\beta\wh P^0$.  Since the three forms of
Hamiltonian Dynamics are
not related by Lorentz transformations, there is no frame in which $\wh P^+=0$.
Therefore, $\dens\neq\exp(-\beta_{LC}\widehat P^-)$
(cf.~Ref.~\cite{Alves,Weldon}).

The quantities $\beta$, $u_\nu$, $\omega$  and $\mu_l$, have the meaning of Lagrange multipliers 
that hold the mean values of the constants of motion fixed, while entropy is maximized.
In an ideal gas for example, the maximum entropy is attained for 
occupation numbers given by Fermi-Dirac and Bose-Einstein statistics \cite{Alves,Beyer},
\beq\label{eq:fermibose}
n(u_\nu p^\nu)=\frac{g}{\euler
^{\beta(u_\nu
p^\nu-\mu)}\pm1}
-\frac{g}{\euler^{\beta(u_\nu p^\nu+\mu)}\pm1},
\eeq
assuming that particles carry charge $+1$ and antiparticles charge $-1$.
(Note that $u_\nu p^\nu\ge0$.)
Here, $p^\nu$ is the 4-momentum of a single particle and 
the degeneracy factor for different spin
states is denoted by $g$. 
The Lagrange multipliers define the equilibrium conditions for two systems.
In complete equilibrium with each other, both systems must have the same values of 
temperature, $u_\nu$, $\omega$  and $\mu_l$, {\em i.e.} no internal motion of 
macroscopic parts of the system is possible in equilibrium (at least in the absence of
vortex lines \cite{Landau5}.) In particular,
the system cannot perform a relative motion
with respect to an external heat bath.
It is important in this respect that the values of these Lagrange 
multipliers do not depend on the total size of the system. 
Thus, even though one can rewrite the statistical operator in terms of 
light-front momentum fractions $x_l$ and the total invariant mass
squared ($M^2$)
as (let $\vec P_\perp=\vec 0_\perp$ and all $Q_l=0$)
\beq
\dens=\exp(-\gamma_x\sum_lx_l-\gamma_{M^2} M^2)
\eeq
the quantities $\gamma_x=u^-P^+/(2T)$ and $\gamma_{M^2}=u^+/(2TP^+)$ 
depend on the overall size of the system and are therefore unsuitable
to define equilibrium conditions. For that reason, some care is required when 
talking about phase diagrams in the $x$-$M^2$ plane, which is often done in the context 
of the Color Glass Condensate \cite{BNL}.

As usual, the (grand) canonical ensemble allows for fluctuations of the total momentum
and of the total charge. Since we have constructed $\dens$ such that it depends 
only on extensive quantities, the relative magnitude of these fluctuations becomes irrelevant for
large systems such as a neutron star, {\em i.e.}
\beq
\frac{\sqrt{\left(\Delta P^\nu\right)^2}}{P^\nu}\propto\frac{1}{\sqrt{N}},
\eeq
for each $\nu$.
Here $N$ is the number of quanta in the system. For a large enough system one can therefore 
use the canonical or the grand canonical ensembles, which are much more convenient 
for calculations than the microcanonical ensemble, even if the values of $P^\nu$, $J_3$ and
$Q_l$ are strictly conserved.
For a large nucleus, $N\gsim 10^3$.
This means a statistical approach is still justified. For a single hadron however, the microcanonical 
ensemble with strict conservation of the constants of motion is more appropriate \cite{liu}.

The simplicity of the light-front vacuum, usually considered an advantage,
seems to bear
problems as far as phase transitions are concerned.  Clearly, thermal field
theory on the
light-front would be rendered useless if it turned out that the zero-mode
problem
\cite{SSB,kazu} would have to be solved before any statements about phase
transitions
could be made. However, the fact that the statistical weight of a
configuration is
maximized for minimal equal-time energy rather than for minimal light-front energy has
important consequences for the formation of condensates at low
temperatures.  The ground
state, {\em i.e.} the state the system is in at $T=0$, is in general
different from the
light-front vacuum. (Indeed, for massive particles $P^-=0$ corresponds to
infinite
$P^0$.)

Alves, Das and Perez \cite{Alves} obtained the self-energy
to one-loop order  in $\phi^4$ theory (${\cal
L}=\partial_\nu\phi^*\partial^\nu\phi-m^2\phi^*\phi-\lambda(\phi^*\phi
)^2$).
Replacing $-m^2$ with $c^2>0$, one immediately obtains spontaneous symmetry
breaking from
their results, including a massless Goldstone boson. No problem arises from 
$1/k^+$-poles.
As usual, one of the fields in this model
develops a non-vanishing expectation value and is shifted correspondingly.
One of the redefined fields, the $\sigma$, acquires a finite mass.
The other field is massless and will be referred to as $\pi$ here.
In the high temperature limit, the masses obtained from the one-loop calculation
of
Ref.~\cite{Alves} are given by the simple expressions 
\beqn m_\sigma^2&=&\left\{
\begin{array}{lr}
2c^2(1-T^2/T_c^2),&\quad T\le T_c\\
\frac{1}{3}\lambda(T^2-T_c^2),&\quad T\ge T_c
\end{array}\right.\\
m_\pi^2&=&\left\{
\begin{array}{lr}
0,&\quad T\le T_c\\
\frac{1}{3}\lambda(T^2-T_c^2),&\quad T\ge T_c
\end{array}\right.,
\eeqn 
with $T_c=3c^2/\lambda$.
These are the standard results \cite{Kapusta}.
The broken phase has one massive sigma and one massless pion.
There is a second order phase-transition at $T=T_c$, where all masses vanish.
In the symmetric phase, all fields have the same mass.

In addition, the authors of Ref.~\cite{Beyer} reproduced the chiral phase
transition in the
Nambu--Jona-Lasinio Model on the light-front.

The possibility of spontaneous symmetry breaking without
zero-modes has
previously been suggested in Ref.~\cite{Thorn} for $\phi^4$(1+1), but these authors did not
study the
finite temperature case. Nevertheless, the picture arising in our work is similar
to that of Ref.~\cite{Thorn}, where $P^-$ is minimized, keeping the total
$P^+$ fixed by
introducing a Lagrange multiplier. In our approach, on the other hand,
the equilibrium configuration
is found by maximizing the entropy for fixed mean
values of all four momentum components.

We conclude that this
approach is poised
for the study of phase transitions in more complicated field theories, such
as QCD.

\section{Fermion Green's function in a medium}
\label{sec:green}

Until now one could get the impression that thermodynamics and statistical
physics on the
light front is identical to the usual instant form approach, except for a
trivial change
of variables. That this is not the case becomes most clear, when one studies
fermions on
the light-front.

In light-front field theory, the Dirac equations can be written as a set of
two coupled
equations for 2-component spinors, see appendix \ref{sec:quant}. Only one
of these
equations contains a time derivative, the other one is a constraint. As a
consequence,
the entire theory can be formulated in terms of 2-component spinors, very
much like a
non-relativistic theory. It is however important to note, that the
2-component theory does
not follow from a local, scalar Lagrangian. The theory is non-local,
because the action
can propagate ``instantaneously" along the light-line, where the
quantization surface
touches the light-cone (see Fig.~\ref{fig:dirac}). This often leads to
difficulties known
as the zero-mode problem. We believe that these difficulties can be overcome by
introducing time-ordered, retarded and advanced Green's functions directly
in the
Hamiltonian formulation. This is done in Ref.~\cite{Landau9} for the
non-relativistic
case, in which non-local operators are common, since particles interact through
potentials. Nevertheless, in a diagrammatic expansion of these Green's
functions, all 4
momentum components are conserved at each vertex. The Green's function is the fundamental 
object of this approach.

To define the various Green's functions of a fermion in a medium, we first
introduce
Heisenberg field operators for the dynamical spinor components ($\alpha\in\{1,2\}$), 
\beqn\label{eq:h1}
\wh \psi_\alpha(r)&=&\euler^{\imath\wh P^-r^+/2}\wh \Psi_\alpha(\ul
r)\euler^{-\imath\wh P^-r^+/2},\\
\label{eq:h2}
\wh \psi_\alpha^\dagger(r)&=&\euler^{\imath\wh P^-r^+/2}\wh
\Psi_\alpha^\dagger(\ul
r)\euler^{-\imath\wh P^-r^+/2}. 
\eeqn 
At equal $r^+$, these Heisenberg
operators fulfill
the same anticommutation relations, Eq.~(\ref{eq:ac}), as the Schr\"odinger
operators.

The time-ordered Green's functions of a fermion in a medium 
is defined in terms of
the dynamical fields only \cite{Alves,Prem},
\beqn\label{eq:green} \imath
G_{\alpha,\beta}(r_1,r_2)&=&\la T_+\wh \psi_\alpha(r_1)\wh
\psi_\beta^\dagger(r_2)\ra\\
&=&
\la\wh \psi_\alpha(r_1)\wh \psi_\beta^\dagger(r_2)\ra\Theta(r_1^+-r_2^+)
-\la\wh \psi_\beta^\dagger(r_2)\wh \psi_\alpha(r_1)\ra\Theta(r_2^+-r_1^+),
\eeqn
where the average, 
\beq
\la\dots\ra=\frac{{\rm Tr}(\dots\dens)}{{\cal Z}},
\eeq 
is to be taken with the appropriate ensemble,
{\em i.e.} canonical or grand canonical for large systems or microcanonical
for hadrons. This definition of the Green's function includes 
the case of zero temperature. Therefore, the conventional light-front quantization
at temperature $T=0$ can be formulated in terms of $G_{\alpha,\beta}$ as well.

We stress that the Green's function in light-front quantization
is not a Lorentz scalar, 
in contrast to equal-time quantization. 
In addition, for isotropic and homogeneous systems, the $G_{\alpha,\beta}$ have the general form
($r=r_1-r_2$)
\beq
G_{\alpha,\beta}(r_1,r_2)=\delta_{\alpha,\beta}G(r),
\eeq
which is completely different from the fermion propagator
in equal-time field theory. While $G$ is an even function of $\vec r_\perp$, 
{\em i.e.}
$G(r^+,r^-,\vec r_\perp)=G(r^+,r^-,-\vec r_\perp)$, it can be seen
 from the definition Eq.~(\ref{eq:green})
that $G$ is not an even function of $r^+$. Neither is $G$ an
even function of $r^-$, because the field operators, Eqs.~(\ref{eq:psi},\ref{eq:psibar}), 
are not defined for negative $k^+$, so that the Fourier transform
$\widetilde G_{\alpha,\beta}(k)$ has a cut
along the negative $k^+$ axis.
This cut is needed to obtain separate information about
fermion and antifermion distributions from the Green's function.
The analytic structure of the light-front Green's function is therefore rather complicated.

The Green's function in momentum space is given by
\beq
\widetilde G_{\alpha,\beta}(k)=\int d^4r \euler^{\imath kr}G_{\alpha,\beta}(r),
\eeq
and the inverse of this transformation is
\beq\label{eq:back}
G_{\alpha,\beta}(r)=\int\frac{d^4k}{(2\pi)^4}\Theta(k^+)\left(\euler^{-\imath k^+r^-/2}+\euler^{+\imath k^+r^-/2}\right)\euler^{-\imath k^-r^+/2}\euler^{+\imath \vec k_\perp\vec r_\perp}\widetilde G_{\alpha,\beta}(k).
\eeq

In a gauge theory, it is also necessary to 
include a (path ordered) gauge connector along the light-cone by redefining the 
fermion fields,
\beqn\label{eq:r1}
\wh \psi_\alpha(r_1)&\to&\wh \psi_\alpha(r_1)
P\exp\{\frac{ig}{2}\int_{-\infty}^{r_1^-} dr^- A^+(r_1^+,r^-,\vec r_{1,\perp})\},\\
\label{eq:r2}
\wh \psi_\beta^\dagger(r_2)&\to&\wh \psi_\beta^\dagger(r_2)
\left[P\exp\{\frac{ig}{2}\int_{-\infty}^{r_2^-} dr^- A^+(r_2^+,r^-,\vec r_{2,\perp})\}
\right]^\dagger.
\eeqn
The path-ordering symbol $P$ arranges the gauge fields $A^+(r)$ in 
order of ascending $r^-$, {\em i.e.} with the largest $r^-$ to the left.
This redefinition does not affect the 
anti-commutation relations Eq.~(\ref{eq:ac}) of the fermion fields. 
Inclusion of the gauge connector  
makes the Green's function gauge invariant in the limit 
$r_1^+-r_2^+\to\pm 0$, $\vec r_{1,\perp}\to\vec r_{2,\perp}$.
The time evolution of $G_{\alpha,\beta}$ is given by
\beq\label{eq:sch}
2\imath\frac{\partial}{\partial r_1^+}G_{\alpha,\beta}(r_1,r_2)
-\la T_+\left[\wh\psi_\alpha(r_1),\wh P^-\right]\wh
\psi_\beta^\dagger(r_2)\ra=
\delta_{\alpha,\beta}
\delta^{(4)}(r_1-r_2).
\eeq
Note that a gauge invariant Hamiltonian $\wh P^-$ alone is not sufficient to
make $G_{\alpha,\beta}$ gauge invariant. One also has to include the 
path-ordered exponentials as in Eqs.~(\ref{eq:r1},\ref{eq:r2}).

A possible subtlety arises in light-cone gauge, $A^+=0$.
Even though the path-ordered exponentials reduce to unity, they may still
influence the time evolution of $G_{\alpha,\beta}$ because their commutator
with $\wh P^-$ will in general not vanish, {\em i.e.}
\beq
\left[\wh\psi_\alpha(r_1)|_{A^+=0},\wh P^-_{A^+=0}\right]
\neq
\left[\wh\psi_\alpha(r_1),\wh P^-\right]_{A^+=0}.
\eeq
In ordinary light-front quantization at $T=0$, the gauge link is 
associated with the single-spin asymmetry \cite{SSA}. Its role at 
finite temperature is yet to be explored.

In addition, the retarded ($R$) and
advanced ($A$) Green's functions are defined by the anticommutators
\beqn \imath
G^{R}_{\alpha,\beta}(r_1,r_2)&=&
\la\left\{
\wh \psi_\alpha(r_1),\wh \psi_\beta^\dagger(r_2)
\right\}\ra
\Theta(r_1^+-r_2^+),\\
\imath G^{A}_{\alpha,\beta}(r_1,r_2)&=&-
\la\left\{
\wh \psi_\alpha(r_1),\wh \psi_\beta^\dagger(r_2)
\right\}\ra
\Theta(r_2^+-r_1^+).
\eeqn 
These Green's functions are needed in finite temperature perturbation theory and
obey Eq.~(\ref{eq:sch}) as well. 
The time-ordered Green's function $G_{\alpha,\beta}$ is related to 
$G_{\alpha,\beta}^A$ and $G_{\alpha,\beta}^R$ by \cite{Landau9},
\beqn
\widetilde G_{\alpha,\beta}(k)=\frac{1}{2}\left(\widetilde G^R_{\alpha,\beta}(k)+\widetilde G^A_{\alpha,\beta}(k)\right)
+\frac{1}{2}{\rm tanh}\left(\frac{uk}{2T}\right)\left(\widetilde G^R_{\alpha,\beta}(k)-\widetilde G^A_{\alpha,\beta}(k)\right).
\eeqn 
In order to obtain this formula, it is important that the Green's function can be
interpreted in terms of transition amplitudes. This is possible, even in a gauge theory,
because the gauge link between the
fermion fields at $r_1$ and $r_2$ can be absorbed into a redefinition
of the field operators, see Eq.~(\ref{eq:r1},\ref{eq:r2}).
Note that at $T=0$, $\widetilde G_{\alpha,\beta}(k)=\widetilde G^R_{\alpha,\beta}(k)$
for positive energy solutions and
$\widetilde G_{\alpha,\beta}(k)=\widetilde G^A_{\alpha,\beta}(k)$
for negative energy solutions.

In the free theory, all Green's functions fulfill the integro-differential
equation, 
\beq
\left[\imath\partial^-_1-\frac{-\partial_{1\perp}^2+m^2}{\imath\partial^+_1}
\right] G^{(0),(X)}_{\alpha,\beta}(r_1,r_2)= \delta_{\alpha,\beta}
\delta^{(4)}(r_1-r_2), 
\eeq 
with boundary conditions appropriate to their
physical
meaning. 
The Green's functions of a fermion in an ideal gas of temperature $T=1/\beta$
were presented first in Ref.~\cite{Alves}. In momentum space, adjusted to  
our notation, they read
\beqn
\widetilde G^{(0)R}_{\alpha,\beta}(k)
&=&\delta_{\alpha,\beta} \frac{k^+}{k^2-m^2+\imath\epsilon\sgn(uk)}, \\
\widetilde G^{(0)A}_{\alpha,\beta}(k)
&=&\delta_{\alpha,\beta} \frac{k^+}{k^2-m^2-\imath\epsilon\sgn(uk)},\\
\widetilde G^{(0)}_{\alpha,\beta}(k)
&=&\delta_{\alpha,\beta} \left({\rm P}
\frac{k^+}{k^2-m^2}
-\imath\sgn(uk)\pi{\rm tanh}\left(\frac{uk}{2T}\right)k^+\delta(k^2-m^2)\right),
\eeqn
where ${\rm P}$ refers to principle value prescription.
In the limit $T\to0$, the light-front Feynman propagator for fermions
turns out to be
\beq
S_F(k)=\delta_{\alpha,\beta} \frac{k^+}{k^2-m^2+\imath\epsilon},
\eeq
which agrees with the result obtained earlier in Ref.~\cite{Prem} with a different technique.

The pole prescriptions $\pm\imath\epsilon\sgn(uk)$ for the retarded and advanced Green's functions
are another manifestation of the special meaning of the equal-time energy. These prescriptions
ensure that $G^{(0)R}_{\alpha,\beta}(r)$ vanishes outside the forward lightcone,
while $G^{(0)A}_{\alpha,\beta}(r)$ is non-vanishing only inside the backward lightcone. That this 
is indeed the case can be seen by comparing $\widetilde G^{(0)R}_{\alpha,\beta}(k)$ 
and $\widetilde G^{(0)A}_{\alpha,\beta}(k)$ to the well-known
propagators of a scalar particle \cite{Peskin},
which differ only by the overall factor $\delta_{\alpha,\beta}k^+$ from the fermion Green's functions
in light-front quantization. 

Most importantly, knowledge of the correct pole prescription eliminates ambiguities 
in the definition of the non-local 
operator $1/k^+$, which appears {\em e.g.} in the free light-cone Hamiltonian,
Eq.~(\ref{eq:pfree}). However, the correct prescription for the $1/k^+$-pole depends on the 
type of Green's function and on the value of the other momentum components. 
The easiest way to overcome these complications is to construct a perturbative expansion of $G$ 
from Eq.~(\ref{eq:sch}). The 4-dimensional $\delta$-function on the right hand side 
leads to 4-dimensional integrals in momentum space, of the kind in Eq.~(\ref{eq:back}). 
This type of perturbation theory is by no means identical to the
usual Feynman perturbation theory in equal-time quantization, as can be seen immediately from the free
propagators. Nevertheless, observable quantities must be the same in both formulations, 
at least within experimental error bars. First applications \cite{Prem}
at $T=0$ indicate that this is the case.

It is possible to introduce the chemical potential in this formalism in a covariant way.
(See also Ref.~\cite{Beyer}.)
The effective Hamiltonian in thermal field
theory on the
light-front is 
\beq \wh{\cal H}=u_\nu\wh P^\nu-\mu\wh Q. 
\eeq 
This operator
propagates
the system in proper time $\tau$. Since in equilibrium $u_\nu$ is
independent of space
time, $r^\nu=u^\nu\tau$. The presence of a chemical potential then
modifies the
generators of space-time translations as if $\mu u^\nu$ were a gauge field. The
modified translation generators read
\beq\label{eq:mu} \trans^\nu=\wh P^\nu-\mu\wh Q u^\nu.
\eeq 
They
propagate the system along trajectories of constant charge and take into account 
that the medium moves as a whole as the 
``test particle''	
represented by the Green's function
propagates from $r_2$ to $r_1$.
The definition of the Heisenberg operators has to be changed accordingly,
{\em i.e.} one needs to replace $\wh P^-$ by $\trans^-$ in Eqs.~(\ref{eq:h1},\ref{eq:h2}). 

Another remarkable property of the light-front Green's functions is, that
if the theory is discretized on a lattice
in $\ul r$-space,
the factor $k^+$ in the numerator
leads to only one pair of fermion doublers. 
Naturally, this simplification comes at a price. In QCD for example, 
the light-front Hamiltonian
contains non-local terms $\propto 1/k^+$ and $\propto 1/(k^+)^2$ \cite{BPPreport}. 
In order to calculate the Green's function on a light-front lattice,
one needs to find field configurations with maximum statistical weight, as
determined from $\dens$. This can in principle be done with the Metropolis algorithm \cite{Metro}.
Unfortunately, each configuration update requires the evaluation of two integrals (because of the
$\propto 1/(k^+)^2$-terms) over the light-cone (for every
lattice site.) It is at present not clear, if this disadvantage outweighs the 
absence of most doubler states. However, we argue that a light-front lattice is advantageous 
at large chemical potential, where Monte-Carlo techniques usually cannot be applied
because of the sign problem.
There is no fermion determinant in the Hamiltonian approach developed here, and also
the chemical potential enters in a different way, see Eq.~(\ref{eq:mu}). A further
investigation of light-front lattice techniques seems therefore interesting to us. Results from transverse lattice calculations at finite temperature have been published recently as well \cite{SD}.

In Discretized Light-Cone Quantization DLCQ \cite{DLCQ}, the alternative non-perturbative 
technique in quantum field theory, it is already known that no fermion doubling problem occurs. 
Furthermore, it is possible to do calculations for a fixed value of the total charge 
without any sign problem. The output of DLCQ is the full spectrum of the theory, {\em i.e.}
all masses of the energy eigenstates and their wavefunctions. 
This is of course much
more information than is needed to calculate bulk thermodynamic properties.
Nevertheless,
from the DLCQ solution of a
field theory, one can directly obtain the partition function.
This was actually done in Ref.~\cite{alex} for QED(1+1). 

Once the Green's function is known numerically from any method, one can calculate all
thermodynamic properties of the system by integrating the relation
\beq
\left(\frac{\partial\Omega}{\partial\mu}\right)_{V,T}=-Q.
\eeq 
In fact, it is sufficient to know $G(r)$ for all $\mu$ and $T$ in the limit $r\to0^-$
to calculate the grand-canonical potential $\Omega$.
From this, the equation of state and hence the phase structure of the theory 
immediately follow.
The relation of $G_{\alpha,\beta}$ to the total charge $Q$ 
will be explained in more detail in the next section.

\section{The light-front density matrix and generalized parton distributions}
\label{sec:gpds}
 
The physical meaning of the time-ordered Green's function
$G_{\alpha,\beta}$ is eludicated further by 
comparing its definition, Eq.~(\ref{eq:green}), with the charge operator in
Eq.~(\ref{eq:q}). One observes
that the net charge distribution in coordinate space is given by 
\beq \label{eq:rel}
Q(\ul
r)=-\imath
G_{\alpha,\alpha}(r^+\to0^-,\ul r|r^+,\ul r). \eeq 
Summation over $\alpha$ is
understood.
The notation $r^+\to0^-$ means that the limit $r^+\to0$ is to be taken
from below. The value of the chemical potential $\mu$ as a function 
of temperature $T$ and total charge
$Q$
is then 
determined by the equation
\beq Q=-\imath\int d^3r G_{\alpha,\alpha}(r^+\to0^-,\ul
r|r^+,\ul r).
\eeq
Of course, the density $Q(\ul r)$ is constant, if 
$G_{\alpha,\beta}(r_1,r_2)=G_{\alpha,\beta}(r_1-r_2)$.

In this section, we introduce the light-front density matrix, which is related
to the light-cone wavefunctions of Ref.~\cite{BL} in the same way as
the one-particle density matrix is related to the wavefunction of a system
in non-relativistic quantum mechanics. If the system is a hadron or a nucleus,
the diagonal elements of the light-front
density matrix are the parton distributions measured in hard scattering experiments, while
the off-diagonal matrix elements turn out to be generalized
parton distributions (GPDs) \cite{gpdf1,Ji,Rady}. This illustrates that the quark and gluon
distributions in a hadron are not classical densities, but quantum mechanical quantities.

Finding all light-front wavefunctions for a hadron is equivalent to
solving the QCD bound state equation
\beq\label{eq:bs}
\wh P^-|h\ra=\frac{\vec P_\perp+M_h^2}{P^+}|h\ra.
\eeq
This has been accomplished for QCD in $1+1$ dimensions \cite{Kent}, but
is a formidable task in any $3+1$ dimensional field theory.
On
the other
hand, if the system is large enough, such as a heavy nucleus, a heavy ion collision, 
or a neutron star, one can hope that a statistical approach may be viable.
In fact, for a system with a very large number of partons, say $>10^3$, 
the density-matrix is probably more useful a concept than the light-front
wavefunctions. 

Knowledge of the density matrix 
enables one to calculate the expectation value of any single-particle operator, 
{\em i.e.} any operator that
can be written as
\beq
\wh F_{\alpha,\beta}(r^+)=\int d^3r\wh\psi^\dagger_\alpha(r)\wh f_{\beta,\gamma}
\wh\psi_\gamma(r)
\eeq
in second quantization.
The density matrix provides only an incomplete description of the system, 
but is sufficient for many applications of statistical physics.

A relation similar to Eq.~(\ref{eq:rel}) exists in condensed matter physics
\cite{Landau9}, but in light-front field theory the situation is more complicated
due to the presence of anti-particles.
For example,
one is tempted to interpret
\beq
Q_{\alpha,\beta}(\ul r_1,\ul r_2)=-\imath G_{\alpha,\beta}(r_2^+\to0^-,\ul r_1|r_2^+,\ul r_2)
=\la\wh \psi_\beta^\dagger(r_2)\wh \psi_\alpha(r_1)\ra \big|_{r_1^+=r_2^+}
\eeq
as the density matrix of the net charge distribution in the system, 
in analogy to non-relativistic Fermi systems. 
However, since the total charge $Q$ can have either sign, one cannot normalize 
$Q_{\alpha,\beta}(\ul r_1,\ul r_2)$ simply by dividing it by $Q$.
Therefore, knowledge
of $Q_{\alpha,\beta}(\ul r_1,\ul r_2)$ does
not immediately enable one to calculate the expectation value
of single-particle operators. One first needs to find 
separate density matrices for fermions and antifermions,
$q_{\alpha,\beta}(\ul r_1,\ul r_2)$ and $\ol q_{\alpha,\beta}(\ul r_1,\ul r_2)$.
Then,
\beq
\la\wh F_{\alpha,\beta}(r^+)\ra=
\frac{\int d^3r_1d^3r_2\wh f^{(1)}_{\beta,\alpha}\left[
q_{\alpha,\beta}(\ul r_1,\ul r_2)+\ol q_{\alpha,\beta}(\ul r_1,\ul r_2)
\right]\delta^{(3)}(\ul r_1-\ul r_2)}
{\int d^3r\left[
q_{\alpha,\alpha}(\ul r,\ul r)+\ol q_{\alpha,\alpha}(\ul r,\ul r)
\right]},
\eeq
with $f^{(1)}_{\beta,\alpha}$ acting on $\ul r_1$.

In the 2-component theory, the separation of fermion and antifermion
distributions requires the evaluation of a Fourier-integral
of the Green's function. 
Since $G_{\alpha,\beta}(r_1,r_2)$ often depends only on the difference
$r=r_1-r_2$, it is natural to introduce the variables
\beqn
R&=&\frac{r_1+r_2}{2},\\
r&=&r_1-r_2.
\eeqn
The first step in finding an expression for the quark density matrix is
to write down the operator structure of $Q_{\alpha,\beta}$,
\beqn\label{eq:wigner}
Q_{\alpha,\beta}(\ul r_1,\ul r_2)&=&
\la\wh \psi_\beta^\dagger(R-\frac{r}{2})
\wh \psi_\alpha(R+\frac{r}{2})\ra \big|_{r_1^+=r_2^+}\\
\nonumber
&=&
\left<
\sum_{\lambda_1,\lambda_2}\int\frac{d^3k_1}{2(2\pi)^3\sqrt{k_1^+}}\Theta(k_1^+)
\int\frac{d^3k_2}{2(2\pi)^3\sqrt{k_2^{+}}}\Theta(k_2^{+})
\right.\\
\nonumber&&\left\{
\wh b^\dagger(\ul k_2,\lambda_2)\wh b(\ul k_1,\lambda_{1})
\chi^\dagger_{\beta}(\lambda_2)\chi_{\alpha}(\lambda_{1})
\euler^{+\imath \ul k_2\cdot\left(\ul R-\frac{\ul r}{2}\right)
-\imath \ul k_{1}\cdot\left(\ul R+\frac{\ul r}{2}\right)}
\right.
\\
\nonumber&+&
\wh d(\ul k_2,\lambda_2)\wh d^\dagger(\ul k_{1},\lambda_{1})
\chi^\dagger_{\beta}(-\lambda_2)\chi_{\alpha}(-\lambda_{1})
\euler^{-\imath \ul k_2\cdot\left(\ul R-\frac{\ul r}{2}\right)
+\imath \ul k_{1}\cdot\left(\ul R+\frac{\ul r}{2}\right)}
\\
\nonumber&+&
\wh b^\dagger(\ul k_2,\lambda_2)\wh d^\dagger(\ul k_{1},\lambda_{1})
\chi^\dagger_{\beta}(\lambda_2)\chi_{\alpha}(-\lambda_{1})
\euler^{+\imath \ul k_2\cdot\left(\ul R-\frac{\ul r}{2}\right)
+\imath \ul k_{1}\cdot\left(\ul R+\frac{\ul r}{2}\right)}
\\
&+&
\left.\left.
\wh d(\ul k_2,\lambda_2)\wh b(\ul k_{1},\lambda_{1})
\chi^\dagger_{\beta}(-\lambda_2)\chi_{\alpha}(\lambda_{1})
\euler^{-\imath \ul k\cdot\left(\ul R-\frac{\ul r}{2}\right)
-\imath \ul k_{1}\cdot\left(\ul R+\frac{\ul r}{2}\right)}
\right\}\right>.
\eeqn
We resort to $A^+=0$ gauge here.
In order to isolate the quark distribution, one needs to Fourier transform over
$r^-$. Since the $k_i^+$ cannot be negative, the $dd^\dagger$-term disappears and
the quark density 
matrix is given by
\beqn\label{eq:quark}\nonumber\lefteqn{
q_{\alpha,\beta}(k^+,\ul R,\vec r_\perp)}\\
&=&\frac{1}{4\pi}
\int dr^-\euler^{+\imath k^+r^-/2}Q_{\alpha,\beta}(\ul r_1,\ul r_2)\\
\nonumber
&=&
\left<
\sum_{\lambda_1,\lambda_2}\int\frac{d^3k_1}{2(2\pi)^3\sqrt{k_1^+}}\Theta(k_1^+)
\int\frac{d^3k_2}{2(2\pi)^3\sqrt{k_2^{+}}}\Theta(k_2^{+})
\right.\\
&&
\nonumber\left\{
\wh b^\dagger(\ul k_2,\lambda_2)\wh b(\ul k_{1},\lambda_{1})
\chi^\dagger_{\beta}(\lambda_2)\chi_{\alpha}(\lambda_{1})
\euler^{+\imath \left(\ul k_2-\ul k_{1}\right)\cdot\ul R
+\frac{\imath}{2}\left(\vec k_{2,\perp}+\vec k_{1,\perp}\right)\cdot\vec r_\perp}
\delta(k^+-\frac{k_1^++k_2^{+}}{2})
\right.
\\
\nonumber
&+&
\wh b^\dagger(\ul k_2,\lambda_2)\wh d^\dagger(\ul k_{1},\lambda_{1})
\chi^\dagger_{\beta}(\lambda_2)\chi_{\alpha}(-\lambda_{1})
\euler^{+\imath \left(\ul k_2+\ul k_{1}\right)\cdot\ul R
+\frac{\imath}{2}\left(\vec k_{2,\perp}-\vec k_{1,\perp}\right)\cdot\vec r_\perp}
\delta(k^+-\frac{k_2^+-k_1^{+}}{2})
\\
&+&
\left.\left.
\wh d(\ul k_2,\lambda_2)\wh b(\ul k_{1},\lambda_{1})
\chi^\dagger_{\beta}(-\lambda_2)\chi_{\alpha}(\lambda_{1})
\euler^{-\imath \left(\ul k_2+\ul k_{1}\right)\cdot\ul R
-\frac{\imath}{2}\left(\vec k_{2,\perp}-\vec k_{1,\perp}\right)\cdot\vec r_\perp}
\delta(k^+-\frac{k_1^{+}-k_2^+}{2})
\right\}\right>.
\eeqn
A similar expression can be obtained for the antiquark density
matrix,
\beq
\ol q_{\alpha,\beta}(k^+,\ul R,\vec r_\perp)
=\frac{1}{4\pi}
\int dr^-\euler^{+\imath k^+r^-/2}\imath G_{\alpha,\beta}(r_2^+\to0^+,\ul r_1,r_2^+,\ul r_2).
\eeq
To obtain the correct order of the creation and annihilation operators for antifermions,
the $r^+\to0$ limit of the Green's function is now taken from the other side.
For definiteness, we shall discuss only the quark density matrix in the following.

The density matrices
$q_{\alpha,\beta}$ (and $\ol q_{\alpha,\beta}$) are related 
to the so-called Wigner function by a
Fourier transform over $r_\perp$. The Wigner function is the 
quantum mechanical analog 
of the classical phase space distribution, to which it reduces in the
limit $\hbar\to 0$. In quantum mechanics, the Wigner function is real 
but not always positive.
It is nevertheless a physical quantity that can be measured in experiment. 
We remark that all properties of the quantum mechanical density matrix, such as 
hermiticity and
positivity of the diagonal matrix elements, also apply to the light-front density 
matrix in $A^+=0$ gauge, since this gauge has only states with positive norm and
no unphysical degrees of freedom.
However, the Wigner function and the light-front density matrix have matrix elements 
that are off-diagonal in Fock-space.
The object defined in Eq.~(\ref{eq:quark}) is similar to the Wigner function 
introduced in Ref.~\cite{wigner}. 

The fermion density matrix contains all information about single quark
properties. It depends on 6 variables and is a $2\times2$ matrix in spinor
space, which can be written as a linear combination of Pauli spin matrices.
The coefficients are the density matrices for
unpolarized ($q$), longitudinal ($\Delta_L q$), and
transverse ($\Delta_T q^{1,2}$) spin distributions,
\beqn\label{eq:decomp}\nonumber
q_{\alpha,\beta}(k^+,\ul R,\vec r_\perp)&=&q(k^+,\ul R,\vec r_\perp)\delta_{\alpha,\beta}
+\Delta_Lq(k^+,\ul R,\vec r_\perp)\sigma_3\\
&+&\Delta_Tq^1(k^+,\ul R,\vec r_\perp)\sigma_1,
+\Delta_Tq^2(k^+,\ul R,\vec r_\perp)\sigma_2,
\eeqn

As already mentioned above,
the diagonal matrix elements in coordinate space
of $q_{\alpha,\beta}(k^+,\ul R,\vec r_\perp)$, {\em i.e.}
the ones with $\vec r_\perp=\vec 0_\perp$, are closely related to 
the usual PDFs \cite{fact}. For instance,
the unpolarized collinear quark density is given by
\beq
q(k^+)=\int d^3Rq(k^+,\ul R,\vec r_\perp=\vec 0_\perp).
\eeq
Strictly speaking, this integral is divergent and needs to be renormalized,
making $q(x)$ scale dependent. At $T=0$, the renormalization group equations
are the QCD evolution equations. A generalization of QCD evolution 
to finite temperature seems possible. In fact, in equal-time quantization,
such a program has been started in Ref.~\cite{wang}. 
Completely analogous expressions exist for $\Delta_{L,T}q(k^+)$.
The parton density is normalized 
such that $\int_0^\infty dk^+q(k^+)=q$, the total 
number of quarks in the system. Note that
\beq
q(k^+)dk^+=\frac{dx}{x}\left<\sum_\lambda\int\frac{d^2k_\perp}{2(2\pi)^3}
b^\dagger(\ul k,\lambda)
b(\ul k,\lambda)\right>
\eeq
depends only on the light-front momentum fraction $x=k^+/P^+$, which goes to
$0$ in the thermodynamic limit. It is therefore appropriate to consider all PDFs as
functions of $k^+$. 

\begin{figure}[t]
  \centerline{\scalebox{1.0}{\includegraphics{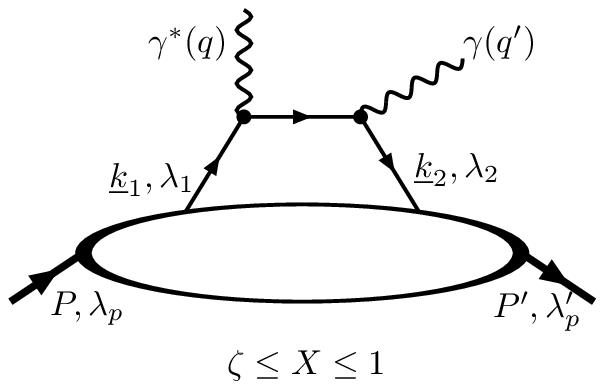}}\hphantom{XXX}
  \scalebox{1.0}{\includegraphics{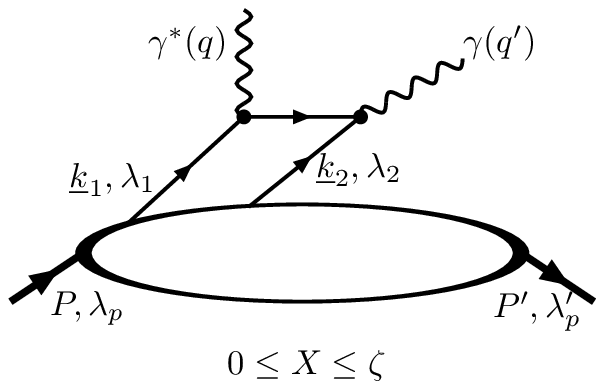}}}
    \center{
    {\caption{\em
      \label{fig:dvcs} The off-diagonal matrix elements of the light-front
      density matrix are related to GPDs, which can be accessed in DVCS.
      The time-ordering shown on the left probes density matrix elements that are off-diagonal
      in momentum space, but diagonal in Fock-space. The other time-ordering (right)
      samples matrix elements that are  off-diagonal in Fock-space.}
    }  }
\end{figure}

The off-diagonal matrix elements 
of
$q_{\alpha,\beta}$ and $\ol q_{\alpha,\beta}$
are related to generalized parton distributions
(GPDs) \cite{gpdf1,Ji,Rady},
which can be accessed {\em e.g.} in deeply virtual Compton scattering (DVCS), 
see Fig.~\ref{fig:dvcs}.
The generalized form factors $H$ and $E$ are defined by (see {\em e.g.} Ref.~\cite{BDH}
for details)
\beqn\label{eq:me} \lefteqn{
\int_{-\infty}^\infty\frac{dr^-}{4\pi}
\euler^{\imath k^+r^-/2}
\la P^\prime,\lambda_{p}^\prime|\psi^\dagger(0)\psi(r)|P,\lambda_{p}\ra\Big|_{r^+=0,\vec r_\perp=\vec 0_\perp}}\\
&=&\frac{1}{2\overline P^+}\overline U(P^\prime,\lambda_{p}^\prime)\left[
H(X,\zeta,t)\gamma^++E(X,\zeta,t)\frac{\imath}{2m_N}\sigma^{+\kappa}(-\Delta_\kappa)\right]
U(P,\lambda_{p})\\
&=&
\frac{\sqrt{1-\zeta}}{1-\frac{\zeta}{2}}H(X,\zeta,t)\delta_{\lambda_{p},\lambda_{p}^\prime}
-\left(\frac{\zeta^2}{4(1-\frac{\zeta}{2})}\delta_{\lambda_{p},\lambda_{p}^\prime}
+\frac{\lambda_{p}\Delta^1+\imath\Delta^2}{2m_N}
\delta_{\lambda_{p},-\lambda_{p}^\prime}\right)\frac{E(X,\zeta,t)}{\sqrt{1-\zeta}},
\eeqn
where $\overline P^\kappa=(P^\kappa+P^{\prime\kappa})/2$. 
The meaning of the four-momenta 
$P$ and $P^\prime$ is illustrated in Fig.~\ref{fig:dvcs}.
The momentum transfer to the target is $\Delta^\kappa=P^\kappa-P^{\prime\kappa}$
and $t=\Delta^2$. We use the kinematic variables of Ref.~\cite{Rady}, {\em i.e.}
$X=k_1^+/P^+$ and $\zeta=1-{P^{\prime+}}/{P^+}$, and employ a frame with $q^+=0$, where $q$
is the four-momentum of the virtual photon.

Since the electromagnetic current touches 
only one quark (at leading twist), the off-forward matrix element
$\la P^\prime,\lambda_{p}^\prime|\psi^\dagger(0)\psi(r)|P,\lambda_{p}\ra$
is diagonal in the momenta ($p_i^+,\vec p_{i,\perp}$) and the spin variables ($\lambda_i$)
of the spectator quarks. Note that
the index $\perp$ refers to the coordinates transverse to the
light-front $z$-direction,
not to the direction of the proton. That way, 
the sum over all $\vec p_{i,\perp}$, including the struck quark,
yields the total transverse momentum of the proton, which in the final state
is different from $\vec 0_\perp$.
Likewise, $\lambda_i$ is the projection 
of the Pauli-Lubansky spin vector onto the light-front $z$-direction 
and therefore different from the quark's helicity.
We remark that the usual light-front variables $x_i$ and $\vec k_{i,\perp}$ of the spectators
(with $\sum \vec k_{i,\perp}=\vec 0_\perp$) are changed in DVCS, but this is a purely
kinematical effect. Therefore, in terms of the variables  $x_i$ and $\vec k_{i,\perp}$,
$\la P^\prime,\lambda_{p}^\prime|\psi^\dagger(0)\psi(r)|P,\lambda_{p}\ra$
can be written as overlap of light-cone wavefunctions \cite{BDH,Diehl2}.
In terms of the actual light-front momenta, which are defined here with respect to the $z$-direction,
the DVCS amplitude is simply the light-front density matrix, {\em i.e.} the sum
($q_{\alpha,\beta}+\ol q_{\alpha,\beta}$),
in momentum space. 

Loosely speaking, the density matrix in coordinate space is related to the
off-forward matrix element (as a function of $\ul k_1$ and $\ul k_2$) by
Fourier transformation. The precise relation can be read off from the phase factors
in Eq.~(\ref{eq:quark}). Because of the positivity constraint on the light-front $+$-momenta,
one has to distinguish four different kinematical domains \cite{Diehl2}.
By inspecting the
$\delta$-functions in Eq.~(\ref{eq:quark}), one finds that for 
$0\le\zeta\le X\le1$, only the $\wh b^\dagger \wh b$-terms contributes to the amplitude (Fig.~\ref{fig:dvcs} (left)),
while for $0\le X\le\zeta\le1$, the electromagnetic current pulls a quark-antiquark  
pair out of the proton ($\wh b\wh d$), see Fig.~\ref{fig:dvcs} (right). 
In the latter region, there is also a contribution
of $\ol q_{\alpha,\beta}$ to the DVCS amplitude. Similar considerations can be made 
for $X,\zeta<0$. This is explained in more detail in Ref.~\cite{Diehl2}.
A complete discussion of the relation between 
$q_{\alpha,\beta}, \ol q_{\alpha,\beta}$ and GPDs is beyond the scope of this paper and will be published elsewhere.
We shall explain only one special case here, which is particularly interesting.

For $\zeta=0$, one has $k^+=k_1^+=k_2^+$, and the GPDs can be identified as 
impact parameter dependent parton densities \cite{Matthias}. In the case
$\lambda_p=\lambda_p^\prime$ we find,
\beqn
q(k^+,\vec b)
&=&\int dR^-\frac{1}{2}\delta_{\beta,\alpha}
q_{\alpha,\beta}(k^+,R^-,\vec b,\vec r_\perp=\vec 0_\perp)\\
&=&
\frac{1}{4\pi k^+}\left<\sum_{\lambda}
\wh b^\dagger(k^+,\vec b,\lambda)\wh b(k^+,\vec b,\lambda)
\right>,\\
q(k^+,\vec b)dk^+
&=&\int \frac{d^2\Delta_\perp}{(2\pi)^2}\euler^{\imath\vec b\cdot\vec\Delta_\perp}H(X,\zeta=0,t).
\eeqn
The creation and destruction operators for a quark at impact parameter $\vec b$
are 
\beqn
\wh b(k^+,\vec b,\lambda)&=&\int\frac{d^2k_{2,\perp}}{(2\pi)^2}\euler^{-\imath\vec k_{2,\perp}\cdot\vec b}b(k^+,\vec k_{2,\perp},\lambda),\\
\wh b^\dagger(k^+,\vec b,\lambda)&=&\int\frac{d^2k_{1,\perp}}{(2\pi)^2}\euler^{+\imath\vec k_{1,\perp}\cdot\vec b}b^\dagger(k^+,\vec k_{1,\perp},\lambda).
\eeqn

The relation between GPDs and $b$-dependent parton densities for $\zeta=0$
was obtained previously by Burkardt
\cite{Matthias} from a different point of view. We stress that for $\zeta\neq 0$,
GPDs are in general not 
probability distributions but density matrices, which do not need to be positive.
Note also that the density matrix only carries information about
single particle properties and contains no information about correlations
between different partons.
Some applications of the Wigner function in the context of DVCS have already been discussed
in Ref.~\cite{wigner}.
The light-front density matrix is a natural extension of
the parton model to quantum mechanics:
classical parton densities are replaced by a density matrix.
It would be interesting to identify other hard processes besides DVCS
which are sensitive
to the quantum mechanical nature of parton distributions.
Most important however is the connection between the density matrix and
the fermion Green's function, because that establishes a common language 
for high energy scattering and statistical QCD.

\section{Summary and Outlook}
\label{sec:summary}

We have investigated the prospects and challenges of
light-front quantization in statistical physics and thermodynamics.
Though most of the results of this
paper apply to any field theory, we are especially interested in QCD.
We have constructed the most general form of the statistical operator $\dens$,
Eq.~(\ref{eq:general}),
using only the light-front Liouville theorem and the Poincar{\'e} 
algebra. Our results generalize earlier findings \cite{diplom,Alves}
to also include rotations and finite density. In particular, we treat the
chemical potential in a covariant way. Remarkably, $\dens$ is not the 
exponential of the light-front Hamiltonian $\wh P^-$, but of the 
operator ${\cal \wh H}=u_\nu\wh P^\nu-\mu\wh Q$, which propagates the system in eigentime
$\tau=u_\nu r^\nu$, {\em i.e.} along its world line.
This is a direct consequence of cluster decomposition, {\em i.e.} of the
requirement that the entropies of two systems separated by a large distance 
are additive. 
Since the spectrum of ${\cal \wh H}$ (for $Q=0$) is that of the equal-time 
Hamiltonian $\wh P^0$ in the rest frame of the system, no problems related to 
light-cone zero-modes arise
in the description of phase-transitions. This finding is also important
for the conventional light-front quantization at zero temperature: at $T=0$,
the system is in the state with the lowest eigenvalue of ${\cal\wh H}$ and not in the
state with lowest $P^-$.

Despite all this, thermal field theory in light-front quantization is not
identical to the usual equal-time formulation. In light-front quantization,
commutation and anticommutation rules are defined on a light-like hypersurface
and only one of the four translation generators, namely $\wh P^-$, depends on the
interaction. The other three terms in $u_\nu\wh P^\nu$ are kinematic.
It is therefore still possible to make use of the light-front 
Fock-state representation at finite temperature and density. The partition function
of a theory for any $T$ and especially any $\mu$
can therefore be obtained from DLCQ (see also Ref.~\cite{alex}). 
The output of DLCQ is the
full spectrum of the theory, eigenvalues and eigenfunctions, which is much more information 
than the single particle properties needed to calculate thermodynamic quantities.
It is therefore our hope to develop a reduced version of DLCQ that yields less
detailed information but is numerically inexpensive enough to be applied
in $3+1$ dimensions. The basis for such a technique would be the light-front density matrix,
which represents the incomplete quantum mechanical description of
a system and replaces the light-front wavefunctions in statistical physics.
The light-front density matrix is closely related to GPDs \cite{wigner}.

Even though formidable numerical challenges still exist in $3+1$ dimensional 
field theories, we believe that this result is important, because
no other first principle technique exists to calculate thermodynamic properties 
at low $T$ and large $\mu$. (See however Ref.~\cite{Fodor} for recent progress 
in lattice QCD with finite chemical potential.) Unlike in lattice gauge theory, there are no problems
arising from dynamical fermions in DLCQ. 
 
The density matrix can be constructed in an elementary way by integrating 
products of light-front wavefunctions over the variables of all but  one
particle. The light-front wavefunctions can be obtained numerically
by diagonalizing the DLCQ Hamiltonian.
This procedure is formally equivalent to taking the limit $r^+\to0^\pm$ of the 
Green's function of a particle in the medium. The latter approach
yields the most obvious generalization of GPDs to finite temperature and density.
It therefore provides a common language for the description of parton distributions 
in high energy scattering and the thermal properties of a fireball.
Although it may be difficult to measure finite temperature PDFs in deep inelastic scattering 
type experiments, these quantities have a profound impact on 
jets passing through the medium \cite{Ivan}
and will also modify the fragmentation 
functions of heavy quarks and other particles \cite{Rainer}. 

We regard the Green's functions 
as the fundamental quantities the theory is built upon, both at $T=0$ or $T\neq0$.
The approach presented in this paper may therefore be regarded as a novel way 
of doing light-front quantization. The advantage of the Green's function formulation is that 
there are no ambiguities in defining the boundary conditions for the 
operator $1/\partial^+$. The pole prescription for the Green's functions are fixed by the
requirement that positive energy solutions have to
propagate into the forward light-cone, while negative energy solutions are to
propagate into the backward light-cone.

The fermion Green's function in the front-form has properties very 
different from the equal-time Green's function. The light-front Green's function is 
a $2\times2$ matrix in spinor space and does not transform as a Lorentz scalar.
Its close relation to GPDs makes light-front field theory appear 
very similar to
condensed matter physics \cite{Landau9}. There are, however, important differences between 
relativistic light-front field theory and nonrelativistic condensed matter physics:
The necessity of antiparticles in a relativistic theory leads to a rather complicated
analyticity structure of light-front Green's functions. In particular, the Green's functions
are not defined along the negative $p^+$-axis. This fact is important to separate particles 
from antiparticles in a theory with 2-component spinors. 

In summary,
we have presented a new formalism for  analyzing relativistic statistical systems based on light-front quantization. The new formalism provides a boost-invariant generalization of thermodynamics, and thus it
has direct applicability to the QCD analysis of heavy ion collisions and other systems of relativistic particles.

\medskip
\noindent {\bf Acknowledgments}: We are grateful to Matthias Burkardt for
making
Ref.~\cite{diplom} available to us.  JR thanks Hans-J\"urgen Pirner for
valuable discussion and 
the Nuclear Physics Group at LBNL for hospitality. JR
acknowledges support from the
Feodor Lynen
Program of the Alexander von Humboldt Foundation.
This work was supported by the U.S.\ Department
of Energy at
SLAC under Contract No.~DE-AC02-76SF00515.

\appendix

\section{Light-Front quantization of the Dirac field}
\label{sec:quant}

The Dirac equations are derived  from the usual fermion Lagrangian ${\cal
L}=\overline{\widetilde\psi}(\imath\fmslash{\partial}-m)\widetilde\psi$ (with
$\overline{\widetilde\psi}=\widetilde\psi^\dagger\gamma^0$, not
$\widetilde\psi^\dagger\gamma^-$). It is a special property of front form
dynamics that
the Dirac equations are a set of two coupled spinor equations
\cite{BPPreport}, only one
of which contains a time derivative, $\partial^-=\partial/\partial
x_-=2\partial/\partial
x^+$. This property of the light-front Dirac equation becomes most
transparent in the
following representation of the $\gamma$-matrices \cite{Hari}. In $2\times
2$ block
matrix notation, 
\beq\label{eq:rep} \gamma^+= \left(\begin{array}{cc}
0&0\\
2\imath&0
\end{array}\right),\quad
\gamma^-= \left(\begin{array}{cc}
0&-2\imath\\
0&0
\end{array}\right),\quad
\vec\gamma_\perp= \left(\begin{array}{cc}
-\imath\vec\sigma_\perp&0\\
0&\imath\vec\sigma_\perp
\end{array}\right),
\eeq where $\vec\sigma_\perp=(\sigma_1,\sigma_2)$ are two of the three
Pauli spin
matrices. The operators projecting out the free ($\Lambda_+$) and the
constrained
($\Lambda_-$) components of a $4$-spinor take the simple form \beq
\Lambda_+=\frac{1}{4}\gamma^-\gamma^+= \left(\begin{array}{cc}
1&0\\
0&0
\end{array}\right),\quad
\Lambda_-=\frac{1}{4}\gamma^+\gamma^-= \left(\begin{array}{cc}
0&0\\
0&1
\end{array}\right),
\eeq and the Dirac equations can be written in terms of 2-component spinors
$\psi$ and
$\eta$, \beqn\label{eq:dyn}
\imath\partial^-\psi(r)&=&\left(-\imath\vec\sigma_\perp\cdot
\vec\partial_\perp-\imath m\right)\eta(r)\\
\label{eq:con}
\imath\partial^+\eta(r)&=&\left(-\imath\vec\sigma_\perp\cdot
\vec\partial_\perp+\imath
m\right)\psi(r). 
\eeqn 
The usual 4-component spinor $\widetilde\psi$ is
related to $\psi$
and $\eta$ by \beq \widetilde\psi(r)= \left(\begin{array}{c}
\psi(r)\\
\eta(r)
\end{array}\right).
\eeq 
The free particle solutions for positive
($\widetilde\psi(r)=u(k)\euler^{-\imath
kx}$) and negative energies ($\widetilde\psi(r)=v(k)\euler^{+\imath kx}$)
read,\beq
u(k,\lambda)=\sqrt{k^+}\left(
\begin{array}{c}
\chi_\lambda \\
\frac{\vec\sigma_\perp\cdot\vec k_\perp+\imath m}{k^+}\chi_\lambda
\end{array}\right)\quad,\quad
v(k,\lambda)=\sqrt{k^+}\left(\begin{array}{c}
\chi_{-\lambda} \\
\frac{\vec\sigma_\perp\cdot\vec k_\perp-\imath m}{k^+}\chi_{-\lambda}
\end{array}
\right),
\eeq
where $\chi_\lambda$ is an eigenstates of $\sigma_3$ with eigenvalue $\lambda$.
These spinors are normalized to
$\bar u(k,\lambda)u(k,\lambda^\prime)=-\bar
v(k,\lambda)v(k,\lambda^\prime)=2m\delta_{\lambda,\lambda^\prime}$. Note also that
$u^\dagger(k,\lambda)u(k,\lambda^\prime)=
v^\dagger(k,\lambda)v(k,\lambda^\prime)=2k^0\delta_{\lambda,\lambda^\prime
}.$ 
We also remark that Eq.~(\ref{eq:rep}) is a chiral representation of the
Dirac algebra, if $\sigma_3$ is chosen diagonal,
\beq
\gamma^5=\frac{\imath}{4!}\epsilon^{\lambda\kappa\nu\rho}\gamma_\lambda
\gamma_\kappa\gamma_\nu\gamma_\rho=
\left(\begin{array}{cc}
\sigma_3&0\\
0&-\sigma_3
\end{array}\right).
\eeq 

In light-front field theory, the initial conditions and the commutation
relations for
field operators are defined on a hypersurface in Minkowski space which is
perpendicular
to a light-like vector $n^\kappa$. The standard choice is
$n^\kappa=(1,0,0,1)$ (here
$\kappa\in\{0\dots3\}$), but it is possible to formulate light-front
quantization in a
manifestly covariant way for general light-like $n^\kappa$ \cite{KC}. In
the following,
we shall use $n^\kappa=(1,0,0,1)$. The 4-velocity $u^\kappa$ then still
needs to satisfy
Weldon's condition $u^\kappa n_\kappa>0$ \cite{Weldon}.

The fermion field operators for the dynamical spinor components
in the Schr\"odinger picture are expanded as
\beqn\label{eq:psi} \wh{\Psi}(\ul r)&=&\sum_\lambda\int
\frac{d^3k}{(2\pi)^3{2\sqrt{k^+}}}\Theta(k^+) \left\{ \wh b(\ul k,\lambda) 
\chi_\lambda 
\euler^{-\imag \ul k\cdot \ul r}+ \wh d^\dagger(\ul
k,\lambda) 
\chi_{-\lambda}
\euler^{+\imag \ul k\cdot \ul r}
\right\},\\
\label{eq:psibar}\wh{\Psi}^\dagger(\ul r)&=&\sum_\lambda\int
\frac{d^3k}{(2\pi)^3{2\sqrt{k^+}}}\Theta(k^+) \left\{ \wh{b}^\dagger(\ul
k,\lambda)
\chi^\dagger_\lambda  
\euler^{+\imag \ul k\cdot \ul r}+ \wh{d}(\ul
k,\lambda)  
\chi^\dagger_{-\lambda}
\euler^{-\imag \ul k\cdot \ul r} \right\}, \eeqn with $\ul
r=(r^-,\vec
r_\perp)$, $\ul k=(k^+,\vec k_\perp)$ and $\ul k\cdot \ul r=k^+r^-/2-\vec
k_\perp\cdot\vec r_\perp$ (see appendix \ref{sec:conventions} for a summary
of the
notations used). The creation and annihilation operators obey the anticommutation
relations \beq \left\{\wh b(\ul k,\lambda), \wh b^\dagger(\ul
k^\prime,\lambda^\prime)\right\} =\left\{\wh d(\ul k,\lambda), \wh
d^\dagger(\ul
k^\prime,\lambda^\prime)\right\} =(2\pi)^32k^+\delta^{(3)}(\ul k-\ul k^\prime)
\delta_{\lambda,\lambda^\prime}, \eeq so that the anticommutator of the
dynamical spinor
components reads \beq\label{eq:ac} \left\{\wh \Psi_\alpha(\ul r),\wh
\Psi_\beta^\dagger(\ul r^\prime) \right\} =\delta_{\alpha,\beta}
\delta^{(3)}(\ul
r-\ul r^\prime). \eeq 
Remarkably, fermions are completely described on the light-front by
2-component spinors.
This gives the theory a certain non-relativistic appearance,
even though all particle and anti-particle states are included.

The formal similarity between light-front field theory and non-relativistic
many-body
physics may occasionally be misleading. For example, the operator of the
conserved charge
is \beqn\label{eq:q} \wh Q&=&\frac{1}{2}\int d^3r
\wh{\overline{\widetilde\Psi}}(\ul
r)\gamma^+\wh{\widetilde\Psi}(\ul r)
=\int d^3r \wh\Psi^\dagger(\ul r)\wh\Psi(\ul r)\\
&=&\sum_\lambda\int\frac{d^3k}{(2\pi)^32k^+}\Theta(k^+) \left[\wh b^\dagger(\ul
k,\lambda)\wh b(\ul k,\lambda)-\wh d^\dagger(\ul k,\lambda)\wh d(\ul
k,\lambda)\right], \eeqn which counts the number of fermions minus
anti-fermions. Here $\wh{\widetilde\Psi}(\ul r)$ is the operator of the usual 4-component spinor field
The
operator $\wh\Psi^\dagger\wh\Psi$ is not positive definite as is the case in
non-relativistic systems. From the anticommutation relations
\beqn
\left\{\wh Q,\wh\Psi(\ul r)\right\}&=&\wh\Psi(\ul r)(2\wh Q-1)=(2\wh Q+1)\wh\Psi(\ul r)\\
\left\{\wh Q,\wh\Psi^\dagger(\ul r)\right\}&=&\wh\Psi^\dagger(\ul r)(2\wh Q+1)=(2\wh Q-1)\wh\Psi^\dagger(\ul r),
\eeqn
one concludes that $\wh\Psi^\dagger(\ul r)$ creates a fermion at position $\ul r$
in coordinate space, while $\wh\Psi(\ul r)$ destroys one at the same point.

Furthermore, we display the expression for the light-front Hamiltonian of
free fermions,
\beqn\label{eq:pfree} \wh P^-&=&\int d^3r
\wh\Psi^\dagger(\ul
r)\frac{-\vec\partial^2_\perp+m^2}{\imath\partial^+}\wh\Psi(\ul r)\\
&=&\sum_\lambda\int\frac{d^3k}{(2\pi)^32k^+}\Theta(k^+) \frac{\vec
k^2_\perp+m^2}{k^+}
\left[\wh b^\dagger(\ul k,\lambda)\wh b(\ul k,\lambda)+\wh d^\dagger(\ul
k,\lambda)\wh
d(\ul k,\lambda)\right]. \eeqn This operator is non-local because of the
$(\imath\partial^+)^{-1}$. Boundary conditions have to be chosen such that
positive
energy particles propagate only into the forward light-cone, while negative
energy
solutions propagate only into the backward light-cone. 

\section{Notation}
\label{sec:conventions}

We follow the conventions of Ref.~\cite{BL}, {\em i.e.} for any vector
$v=(v^+,v^-,\vec
v_\perp)$, we define $v^+=v^0+v^3$ and $v^-=v^0-v^3$. The metric tensor is then
off-diagonal \cite{BPPreport}, so that the scalar product is \beq
v_1v_2=\frac{v_1^+v_2^-}{2}+\frac{v_1^-v_2^+}{2}-\vec v_{1\perp}\cdot\vec
v_{2\perp}=v_{1+}v_2^++v_{1-}v_2^--\vec v_{1\perp}\cdot\vec v_{2\perp}.
\eeq Partial
derivatives refer to coordinate space, \beqn
\partial^-&=&\frac{\partial}{\partial r_-}=2\frac{\partial}{\partial r^+},\\
\partial^+&=&\frac{\partial}{\partial r_+}=2\frac{\partial}{\partial r^-}.
\eeqn The 4-dimensional volume element is \beq
d^4r=\frac{1}{2}dr^+dr^-d^2r_\perp, \eeq
and we denote the 4-dimensional $\delta$-function by \beq
\delta^{(4)}(r)=2\delta(r^+)\delta(r^-)\delta(\vec
r_\perp)=\delta(r_-)\delta(r^-)\delta(\vec
r_\perp)=\delta(r^+)\delta(r_+)\delta(\vec
r_\perp), \eeq so that $\int d^4r\delta^{(4)}(r)=1$.

In addition, it is useful to introduce a shorthand notation for spacelike
vector-components, where spacelike means the three components $\ul
r=(r^-,\vec r_\perp)$
of coordinate space vectors, but $\ul k=(k^+,\vec
k_\perp)$ for
momenta. We then define \beqn
\ul r\cdot\ul k&=&\frac{r^-k^+}{2}-\vec r_\perp\cdot\vec k_\perp,\\
d^3r&=&dr^-d^2r_\perp,\\
d^3k&=&dk^+d^2k_\perp,\\
\delta^{(3)}(r)&=&\delta(r^-)\delta(\vec r_\perp)
=\frac{1}{2}\int \frac{d^3k}{(2\pi)^3}\Theta(k^+)\left(\euler^{-\imath\ul
r\cdot\ul k}+\euler^{+\imath\ul
r\cdot\ul k}\right),\\
\delta^{(3)}(k)&=&\delta(k^+)\delta(\vec k_\perp)
=\frac{1}{2}\int \frac{d^3r}{(2\pi)^3}\euler^{\imath\ul
r\cdot\ul k}. \eeqn 
The Lorentz invariant phase space element is given by
\beq
\frac{d^4k}{(2\pi)^4}2\pi\delta^+(k^2-m^2)=\frac{d^3k}{(2\pi)^32k^+}\Theta(k^+)
\eeq
Furthermore, we denote the step-function by
\beq\label{eq:step} \Theta(c)=\left\{
\begin{array}{cr}
1 & {\rm for}\quad c>0 \\
0  & {\rm for}\quad c<0.
\end{array}\right.
\eeq and the sign-function by \beq\label{eq:sgn} \sgn(c)=\left\{
\begin{array}{cr}
+1 & {\rm for}\quad c>0 \\
0  & {\rm for}\quad c=0\\
-1  & {\rm for}\quad c<0.
\end{array}\right.
\eeq


\end{document}